# A Combined Theoretical and Experimental Study of the Phase Coexistence and Morphotropic Boundaries in Ferroelectric-Antiferroelectric-Antiferrodistortive Multiferroics

Anna N. Morozovska[1*], Dmitry V. Karpinsky[2a,b†], Denis O. Alikin[3,4], Alexander Abramov[4], Eugene A. Eliseev[5], Maya D. Glinchuk[5], Andrii D. Yaremkevich[1], Olena M. Fesenko[1], Tamara V. Tsebrienko[1], Andrius Pakalniškis[6], Aivaras Kareiva[6], Maxim V. Silibin[7a,b], Vitali V. Sidski[8], Sergei V. Kalinin[9‡], and Andrei L Kholkin[3,4*,10§],

[1] Institute of Physics, National Academy of Sciences of Ukraine, 46, pr. Nauky, 03028 Kyiv, Ukraine

[2a] Scientific-Practical Materials Research Centre of NAS of Belarus, 220072 Minsk, Belarus

[2b] South Ural State University, 454080, Chelyabinsk, Lenin av., 76, Russia

[3] Department of Physics & CICECO-Aveiro Institute of Materials, 3810-193, University of Aveiro, Portugal

[4] School of Natural Sciences and Mathematics, Ural Federal University, Russia

[4*] National University of Science and Technology MISiS, 119049, Moscow, Russia

[5] Institute for Problems of Materials Science, National Academy of Sciences of Ukraine, Krjijanovskogo 3, 03142 Kyiv, Ukraine

[6] Institute of Chemistry, Vilnius University, Naugarduko 24, LT-03225 Vilnius, Lithuania

[7a] National Research University of Electronic Technology, 124498 Zelenograd, Moscow, Russia

[7b] Institute for Bionic Technologies and Engineering, I.M. Sechenov First Moscow State Medical University, 119991 Moscow, Russia

[8] Gomel State University of Belarus, Gomel, Belarus

[9] Center for Nanophase Materials Sciences, Oak Ridge National Laboratory, Oak Ridge, TN 37831

[10] ITMO University, St. Petersburg 197101, Russia.

---

[*] Corresponding author, e-mail: anna.n.morozovska@gmail.com

[†] Corresponding author, e-mail: dmitry.karpinsky@gmail.com

[‡] Corresponding author, e-mail: sergei2@ornl.gov

[§] Corresponding author, e-mail: kholkin@gmail.com

**Abstract**

The physical nature of the ferroelectric (FE), ferrielectric (FEI) and antiferroelectric (AFE) phases, their coexistence and spatial distributions underpin the functionality of antiferrodistortive (AFD) multiferroics in the vicinity of morphotropic phase transitions. Using Landau-Ginzburg-Devonshire (LGD) phenomenology and a semi-microscopic four sublattice model (FSM), we explore the behavior of different AFE, FEI and FE long-range orderings and their coexistence at the morphotropic phase boundaries in FE-AFE-AFD multiferroics. These theoretical predictions are compared with the experimental observations for dense $Bi_{1-y}R_yFeO_3$ ceramics, where $R$ is Sm or La atoms with the fraction $0 \leq y \leq 0.25$, as confirmed by the X-ray diffraction (XRD) and Piezoresponse Force Microscopy (PFM). These complementary measurements were used to study the macroscopic and nanoscopic transformation of the crystal structure with the doping. The comparison of the measured and calculated AFE/FE phase fractions demonstrate that the LGD-FSM approach well describes the experimental results obtained by XRD and PFM for $Bi_{1-y}R_yFeO_3$. Hence, this combined theoretical and experimental approach provides further insight into the origin of the morphotropic boundaries and coexisting FE and AFE states in model rare-earth doped multiferroics.

## I. INTRODUCTION

Multiferroic materials with coupled ferroelectric (**FE**), ferrielectric (**FEI**) or antiferroelectric (**AFE**), and ferromagnetic (**FM**) or antiferromagnetic (**AFM**), and antiferrodistortive (**AFD**) long-range ordering remain at the forefront of modern materials science research. This interest is due both to the broad variety of current and potential applications and the insight into fundamental physics of multiferroics [1, 2, 3, 4, 5]. Applications such as ferroelectric tunneling barriers, light-assisted ferroic dynamics, spin-driven effects, and ultrafast magnetoelectric switching for memory applications have been demonstrated [6, 7, 8].

Among the multiferroic materials, particularly of interest are bulk and nanosized $BiFeO_3$ (**BFO**) extensively studied over the last decade due to its multiferroic properties [9, 10, 11, 12, 13, 14]. Advances in applications drive the need for fundamental understanding of FE, AFE and AFD order parameter dynamics in the materials, which, in turn, necessitates the study of mesoscale phenomena such as local polarization switching, topological defects, and atomic scale phenomena at surfaces and interfaces [15, 16, 17, 18]. On the nanometer scale, the breakthrough in understanding the functional properties of ferroics and multiferroics have been achieved via Scanning Probe Microscopy studies, such as Piezoresponse Force Microscopy (**PFM**) [19, 20], and aberration corrected High-Resolution Scanning Transmission Electron Microscopy (**HR STEM**) [21, 22, 23, 24]. For BFO, these have been used to reveal complex nanoscale evolution of domain structure [25, 26, 27, 28, 29, 30], including vortices and vertices [31, 32, 33, 34, 35]. Domain walls and mesoscale structural defects in BFO were found to exhibit unusual electrophysical properties, such as domain-wall mediated conduction and enhanced magneto-transport [20, 21, 32, 36, 37, 38, 39, 40, 41].

While for pure BFO these studies provide a high veracity insight into polarization dynamics and domain structures, the situation becomes much more complex and unexplored for the rare-earth doped $Bi_{1-y}R_yFeO_3$, $0 \leq y < 1$, where the **rare-earth "$R$"** is Sm, La, Pr and other atoms. In this case, in addition to the pure-phase AFD-FE ordering and canted antiferromagnetic subsystem, additional symmetry lowering, spatial modulation, and order parameters can emerge [42, 43, 44, 45, 46, 47]. Rare-earth doped BFO exhibits the structural state, where three phases coexist: ferroelectric rhombohedral *R3c* phase (**R-phase**), antiferroelectric antipolar orthorhombic *Pbam* or *Pnam* phase (**O-phases**), and non-ferroelectric nonpolar *Pnma/Pbnm* phase [48, 49, 50]. In dependence on the dopant nature, the phase boundaries are more pronounced or diffuse [48, 49, 50, 51]. $Bi_{1-y}Sm_yFeO_3$ and $Bi_{1-y}La_yFeO_3$ solid solutions are known to possess gradual phase transition with the increase of the dopant concentration from the prototypical ferroelectric R-phase of pure $BiFeO_3$ to an O-phase at 20% Sm and La, respectively [51].

Recently, spatially resolved studies by PFM revealed a strong change of the domain structure and polar phase distribution with doping: the transformation of the micro-scale lamellar domain structure to curved domains [52, 53]. According to HR-STEM measurements, the transition is also readily observable in maps of the polar atomic displacement $P$ between the A-site and B-site cation sublattices [51]. For the pure $BiFeO_3$ $P$ is a proxy for the electrical dipole moment and its polydomain distribution is a characteristic of a rhombohedral phase. The distribution of $P$ in $Bi_{0.8}Sm_{0.2}FeO_3$ reveals the large oscillation of its in-plane component corresponding to the AFD orthorhombic structure. The intermediate composition $Bi_{0.93}Sm_{0.07}FeO_3$ exhibits a mixed-structure, with the small domains identifiable to both structures appear in the near interface region (see fig. 1 in Ref.[ 51]).

The coexistence of the structural states was also highly dependent on the conditions of the ceramic synthesis: mechanochemical activation leads to the characteristic scale decrease of the polar phase inclusions and to the appearance of the vortex-like polarization states [35]. At the same time, the domain walls and phase boundaries are highly mobile in the morphotropic (**MP**) phase region and the phase coexistence can be significantly modified by the application of the external electric field, suggesting the electric field-induced phase transition [52, 54]. Despite the large amount of experimental data, a theoretical analysis of the AFE-FE phase coexistence in the BFO-based systems is not performed to the best of our knowledge.

A four sub-lattices model (**FSM**) [55] has been developed for the analytical description of the corresponding A-cation sublattice displacements in $Bi_{1-y}R_yFeO_3$. However, the FSM itself cannot provide a link between the "additional order parameters" – atomic displacements $U$ of A-cations (which are $R$ or Bi) in a BiFeO$_3$ structure, and "intrinsic" long-range parameters, such as FE polarization **P**, antipolar parameter **A**, and AFD order parameter **Φ**, that can be the out-of-phase static rotations of some atomic groups, e.g. oxygen octahedrons. The interplay between the order parameters rules the phenomena taking place at the domain walls, surfaces, and interfaces of AFE-FE-AFD multiferroics. Complementary to FSM, Landau-Ginzburg-Devonshire (**LGD**) approach allows the description of **A**, **P** and **Φ** dynamics on multiple length scales [12, 13, 18], but the corresponding coupling terms cannot be determined from mesoscopic models and necessitate atomistic studies. Recently we have developed LGD theory combined with FSM and establish the primary driving forces for a formation of spatially-modulated phases (**SMPs**) in the rare-earth doped orthoferrites [56]. We have shown that the hybrid FSM-LGD approach can provide unique link between the atomic displacements **U** of R/Bi cations, FE, AFE, and AFD long-range parameters.

In this work, the origin and behavior of various-type AFE, FE and FEI long-range orderings, their coexistence and emerging MP boundaries in rare-earth doped ferrites are discussed. The quantitative results of the phase analysis in the Sm and La doped BFO done by XRD and PFM is analyzed based on LGD-FSM approach. Bi$_{1-y}$R$_y$FeO$_3$, where the rare-earth element *R* is Sm or La atoms, which fraction *y* varies in the range $0 \leq y < 0.25$. The solid solutions obtained by sol-gel method and solid-state synthesis are studied and compared, which allows to elucidate a contribution from the dopant type and concentration on the phase composition of the morphotropic states.

## II. THEORETICAL APPROACH

LGD functional of an AFE-FE-AFD multiferroic utilizes Landau-type power expansion, $G_L$, that includes the contributions of FE and AFE polarization components $P_i$ and $A_i$, and the contribution of AFD order parameter $\Phi_i$; the biquadratic couplings between all order parameters, as well their gradient energy $G_{grad}$, electrostatic, elastic and electrostriction coupling energy, $G_{els}$. Thus, the bulk density of the functional $G_{LGD}$ is [56]:

$$G_{LGD} = G_L + G_{grad} + G_{els}. \tag{1a}$$

Electrostatic and flexoelectric contributions are neglected in this work, since we consider uncharged domain structures only, and the role of flexoelectric coupling has been studied earlier [55, 56]. The energies $G_{grad}$ and $G_{els}$ are listed in Ref.[56].

The Landau energy $G_L$ includes FE, AFE, AFD energies and their biquadratic coupling energy:

$$G_L = G_\Phi + G_P + G_A + G_{PA} + G_{\Phi A} + G_{P\Phi}, \tag{1b}$$

The energies $G_\Phi$, $G_{\Phi A}$ and $G_{P\Phi}$ are listed in Ref.[56], FE, AFE and AFE-FE biquadratic coupling energies are

$$G_P = a_i(T)P_i^2 + a_{ij}P_i^2 P_j^2 + a_{ijk}P_i^2 P_j^2 P_k^2, \qquad G_A = c_i(T)A_i^2 + c_{ij}A_i^2 A_j^2 + c_{ijk}A_i^2 A_j^2 A_k^2, \qquad (2a)$$

$$G_{PA} = \chi_{ijkl}P_i P_j A_k A_l. \qquad (2b)$$

Here the summation is employed over repeated indexes. The coefficients $a_i$ and $c_i$ are temperature dependent, $a_i = a_T(T - T_C)$, $c_i = c_T(T - T_A)$, where $T_C$ and $T_A$ are the temperatures of FE and AFE phases absolute instability, respectively. Following Ref. [56], we regard that the strength of FE-AFE coupling tensor $\chi_{ijkl}$ is defined by the dimensionless temperature-independent scalar parameter $\chi$.

For the sake of simplicity, we assume that the temperatures $T_C$ and $T_A$ are dependent on the concentration $y(\mathbf{r})$ of the rare-earth element $R$ and its spatial variations $\delta y(\mathbf{r})$. Below we consider the functional dependence of the temperatures $T_C$ and $T_A$ on $y$, $T_C = T_{C0}(1 - C(y))$ and $T_A = T_{A0}(1 - D(y))$, where the functions $C(y)$ and $D(y)$ are analytical and continuous. All other coefficients in Eqs.(2a)-(2b) are regarded concentration-independent [56].

Note that the FE-AFE transition takes place with $R$ concentration increase up to the critical value $y_0$ at the fixed temperature (e.g. room) and all other the same conditions. This fact causes the equalities of FE and AFE phases Curie temperatures, $T_{A0} = T_{C0} = T_0$, equalities of the functions $D(y) = -C(y - y_0)$, and coefficients $a_{ij} = c_{ij}$, and $a_{ijk} = c_{ijk}$. The equalities are in a complete agreement with Kittel-type models of equivalent dipole sublattices. The assumption allows to introduce the temperature-dependent dimensionless doping factor $\epsilon$:

$$\epsilon(y) = \frac{T_0}{T_0 - T}C(y - y_0), \qquad (3a)$$

where the concentration $y_0$ corresponds to the MP boundary, and $C(y - y_0)$ is an analytical continuous function. For a BFO:R the value $y_0$ varies in the range $(10 - 45)\%$ [44-46]. Expression (3a) allows us to express the coefficients $a_i$ and $c_i$ through it the doping factor [56]:

$$a_i = a_0(1 - \epsilon) \text{ and } c_i = a_0(1 + \epsilon), \text{ where } a_0 = a_T(T_0 - T). \qquad (3b)$$

Next, using Dzyaloshinsky substitution [57], we relate the polarization $\mathbf{P}$ and three antipolar ($\mathbf{A}$, $\mathbf{B}$ and $\widetilde{\mathbf{A}}$) order parameters with the four atomic displacements $U_i^{(m)}$ from the centrosymmetric positions of A-cations in an ABO$_3$ structure [55, 56]:

$$P_i = \frac{q}{2}\left(U_i^{(1)} + U_i^{(2)} + U_i^{(3)} + U_i^{(4)}\right), \quad A_i = \frac{q}{2}\left(U_i^{(1)} - U_i^{(2)} - U_i^{(3)} + U_i^{(4)}\right), \qquad (4a)$$

$$B_i = \frac{q}{2}\left(U_i^{(1)} - U_i^{(2)} + U_i^{(3)} - U_i^{(4)}\right), \quad \tilde{A}_i = \frac{q}{2}\left(U_i^{(1)} + U_i^{(2)} - U_i^{(3)} - U_i^{(4)}\right). \qquad (4b)$$

Here $q \cong \frac{Q^*}{a^3}$ is a dimension factor proportional to the effective Born charge $Q^*$ divided by the cubic unit cell volume $a^3$ (corresponding to a cubic m3m parent phase). The superscript $m$=1, 2, 3, 4 enumerates the FSM displacement vectors $\mathbf{U}$, which correspond to one of the four $R$/Bi sublattices in the Bi$_{1-y}R_y$FeO$_3$; and the subscript $i$=1, 2, 3 enumerates the components of the vectors $U_i^{(m)}$ in the $m$-th sublattice.

Landau expansion of the free energy for an AFD multiferroic with m3m-symmetry parent phase contains quadratic and bilinear contributions of the quantities $\mathbf{U}^{(m)}$, and thus has the form

$$G_L\left[\mathbf{U}^{(i)}\right] = \mu_{ij}\left(\mathbf{U}^{(i)}, \mathbf{U}^{(j)}\right) + \eta_{ijkl}\left(\mathbf{U}^{(i)}, \mathbf{U}^{(j)}\right)\left(\mathbf{U}^{(k)}, \mathbf{U}^{(l)}\right), \qquad (5)$$

here $\left(\mathbf{U}^{(\xi)}, \mathbf{U}^{(\zeta)}\right) = \sum_i U_i^{(\xi)} U_i^{(\zeta)}$ is the scalar product of the corresponding vectors. For derivation of Eq.(5) and link of the coefficients $\mu_{ij}$ and $\eta_{ijkl}$ with LGD-expansion coefficients see Appendix A from Ref. [56].

For the simplest model case, when the gradient energy of the spatial inhomogeneities can be neglected, the bulk part of Landau free energy density (2) as a function of the dimensionless polar and antipolar order parameters magnitudes, $p$ and $a$, is [56]:

$$g_L = -(1-\epsilon)\frac{p^2}{2} - (1+\epsilon)\frac{a^2}{2} + \frac{p^4 + a^4}{4} + \frac{\chi}{2}p^2 a^2, \qquad (6)$$

note that the condition $\chi > -1$ imposed on the AFE-FE coupling strength $\chi$ should be fulfilled for the stability of the functional (6). Also, we will regard that $|\epsilon| < 1$. Note that the y-dependent doping factor $\epsilon(y)$ and the y-independent FE-AFE coupling strength $\chi$, can control the mutual asymmetry of the four sublattices [56]. Note, that only $p$-values can be observable by PFM, while $a$-values can be observable by XRD.

## III. EXPERIMENTAL DETAILS

The samples for the experimental study of the phase coexistence were prepared by the sol-gel method, Bi$_{1-y}$Sm$_y$FeO$_3$ ceramics, and by solid state sintering La-doped BFO ceramics [58]. For PFM measurements the samples surface was prepared by a step-by-step polishing with the abrasive down to 250 nm diamond paste. At the final stage, the surface was mechanochemically treated by the 60 nm colloidal silica solution. Further samples were heated up to 600 °C to remove mechanical stress after polishing. Then samples were cleaned in acetone and glued to the conductive metal discs.

X-ray diffraction (XRD) measurements were performed using Rigaku diffractometer using Cu Kα radiation in the 2Theta range 10º - 80º with a step of 0.02º. PFM mode of the scanning probe microscope MFP-3D (Asylum Research, Oxford Instruments, Abingdon, UK) was used to visualize domain structure and inspect local piezoelectric response distribution across the surface. Measurements were done with tungsten carbide-coated HA_NC (Scansens, Hamburg, Germany) probes (30 nm nominal tip radius, 900 kHz first flexural contact resonance frequency, and 12 N/m spring constant). AC voltage (20 kHz, 5 Vrms) was applied to the tip. Both vertical and lateral PFM signals were recorded. PFM phase offset was corrected by the rotation at phase angle and subtraction of the frequently-dependent background according to ref. [59]. The corrected images of R·cosθ PFM signal are further referred to as the piezoresponse or PFM images. Quantification of the piezoresponse was done with vertical and lateral force curve measurements [60, 61]. The separation of the piezoelectrically-active and piezoelectrically-inactive phases was done based on the procedure described earlier in ref. [50]. The fractions of the phases were extracted as a percentage of the area with piezoelectrically-active and inactive phases in the scan. Multiple scans were used for the calculation of the average fractions.

## IV. PHASE DIAGRAM AND LONG-RANGE ORDER PARAMETERS

The hybrid-type LGD-FSM free energy, described by Eqs. (1)-(6), is characterized by several long-range ordered AFD phases, listed in **Table I**. Since all these phases are AFD, we omit the abbreviation in the table for the sake of brevity. The cations displacements $U$ and related parameters $P, A, B$ and $\widetilde{A}$ are shown by arrows in the second column of **Table I**. The second column of the table reflects the important feature of the modified FSM – it is a spatial 3D model, since $U^{(m)}$ is a 3D vector, and the third column contains the information about the vector magnitudes only.

For zero spatial gradients, minimization of the LGD-FSM free energy density allows a rhombohedral ferroelectric **FE** R-phase, an orthorhombic antiferroelectric **AFE** O-phase, their spatial coexistence in the **AFE-FE state**, and a true mixed ferrielectric **FEI** phase. The spontaneous values of the order parameters, corresponding free energy, and these phases stability conditions of are summarized in the third and fourth columns of **Table I**.

**Table I.** Long-range ordered AFD phases in LGD-FSM model and necessary conditions of their stability. Adapted and accomplished in comparison with Ref. [56].

| Phase acronym and description | Atomic displacements $U^{(1)}, U^{(2)}, U^{(3)}, U^{(4)}$ and related parameters $P, A, B$ and $\widetilde{A}$ | Magnitude of dimensionless order parameters $p$ and $a$ and Landau energy | Condition of the phase absolute stability or metastabilty |
|---|---|---|---|
| **FE phase** (ferro-electric R-phase) | $U^{(1)} = U^{(2)} = U^{(3)} = U^{(4)}$ $P \neq 0, A = 0, B = 0, \widetilde{A} = 0$ Polar displacements have the same direction, but may differ in length within one domain | $p = \pm\sqrt{1-\epsilon}, a = 0,$ $G_{FE} = -\frac{1}{4}(1-\epsilon)^2$ | Absolutely stable when $0 < \epsilon$ and $\chi > \chi_{cr}^p(\epsilon)$, where $\chi_{cr}^p(\epsilon) = \frac{1+\epsilon}{1-\epsilon}$.* |
| **AFE phase** (antiferroelectric O-phase) | $A = A_S$, (or $\tilde{A} = A_S$) $P = 0$, $B = 0$, $\tilde{A} = 0$ (or $A = 0$) $U^{(1)} = U^{(2)} = -U^{(3)} = -U^{(4)}$ alternatively $P = 0$, $A = 0$, $\tilde{A} = 0$, $B = B_S$, $U^{(1)} = -U^{(2)} = U^{(3)} = -U^{(4)}$ Two equivalent sublattices with opposite directions of polar displacements | $p = 0$, $a = \pm\sqrt{1+\epsilon}$, $G_{AFE} = -\frac{1}{4}(1+\epsilon)^2$ | Absolutely stable when $\epsilon > 0$ and $\chi > \chi_{cr}^a(\epsilon)$, where $\chi_{cr}^a(\epsilon) = \frac{1-\epsilon}{1+\epsilon}$. |
| **FEI phase** (ferrielectric phase, O/R mixture) | $P \neq 0$, $A \neq 0$, (or $\tilde{A} \neq 0$) $B = 0$, $\tilde{A} = 0$ (or $A = 0$) $U^{(1)} = U^{(2)} = U_P$, $U^{(3)} = U^{(4)} = U_A$ Both polar and anti-polar displacements are present in quadrupled unit cell with four sublattices | $p = \pm\sqrt{\frac{1-\epsilon-\chi(1+\epsilon)}{1-\chi^2}}$, $a = \pm\sqrt{\frac{1+\epsilon-\chi(1-\epsilon)}{1-\chi^2}}$ $G_{FEI} = -\frac{1+\epsilon^2-\chi(1-\epsilon^2)}{2(1-\chi^2)}$ | stable when $\chi < \frac{1-|\epsilon|}{1+|\epsilon|}$ and $|\epsilon| < 1$ |
| **AFE-FE state** (mixed state with coexisting FE and AFE regions) | Spatial regions with polar and anti-polar displacements alternate in 3D-space. Allowing for the gradient effects the state can be incommensurately spatially modulated. The boundary between coexisting AFE and FE states is morphotropic and/or diffuse (see **Fig. 3a**) | FE regions with $p = \pm\sqrt{1-\epsilon}$, $a = 0$ alternate with AFE regions with $p = 0$, $a = \pm\sqrt{1+\epsilon}$ $G_{AFE} < G_{FE}$ at $\epsilon > 0$ and $G_{FE} < G_{AFE}$ at $\epsilon < 0$ | metastable when $\chi > \frac{1+|\epsilon|}{1-|\epsilon|}$ and $|\epsilon| < 1$ |

*The critical values of the FE-AFE coupling constants, $\chi_{cr}^p(\epsilon) = \frac{1+\epsilon}{1-\epsilon}$ and $\chi_{cr}^a(\epsilon) = \frac{1-\epsilon}{1+\epsilon}$, corresponding to the FEI-FE and FEI-AFE phase boundaries, are derived in **Appendix A**, Supplementary Materials.

The free energy (6) as a function of polar and antipolar order parameters, $p$ and $a$, is shown in **Fig. 1** for different $\epsilon$ and $\chi$ values. From **Fig. 1** one can see that, for $\chi < 0$ and small $\epsilon$ values (as well as for $\epsilon = 0$ and $\chi \neq 1$) the free energy contains four potential wells of the same depth, separated by the four saddles and a central maximum. The case $\chi < 0$ corresponds to FEI phase, since both order parameters are nonzero, namely their magnitudes are $p = \pm\sqrt{\frac{1-\epsilon-\chi(1+\epsilon)}{1-\chi^2}}$ and $a = \pm\sqrt{\frac{1+\epsilon-\chi(1-\epsilon)}{1-\chi^2}}$. The depth of the four potential wells is different for $\chi > 1$ and small negative $\epsilon < 0$; there are two deeper wells corresponding to $p = \pm\sqrt{1-\epsilon}$ and $a = 0$, and two shallower wells corresponding to $a = \pm\sqrt{1+\epsilon}$ and $p = 0$. Hence, the case $\chi > 1$ and small $\epsilon < 0$ corresponds to the mixed FE-AFE state, where FE regions are stable and AFE regions are metastable. An almost round potential well indicating on a FEI→AFE-FE transition is realized in a special case $\chi = 1$ and $|\epsilon| \ll 1$. The well becomes exactly round at $\chi = 1$ and $\epsilon = 0$, since the equilibrium values of the order parameters are given by equation $p^2 + a^2 = 1$ in the case. For the case $\chi > 1$ and small positive $\epsilon > 0$ the two deeper wells correspond to $a = \pm\sqrt{1+\epsilon}$ and $p = 0$, while two shallower wells correspond to $p = \pm\sqrt{1-\epsilon}$ and $a = 0$. Hence, the case $\chi > 1$ and small $\epsilon < 0$ corresponds to the mixed FE-AFE state, where AFE regions are stable and FE regions are metastable. Note that the free energy relief transforms into a thin cross for the case $\chi \gg 1$ and $|\epsilon| \ll 1$.

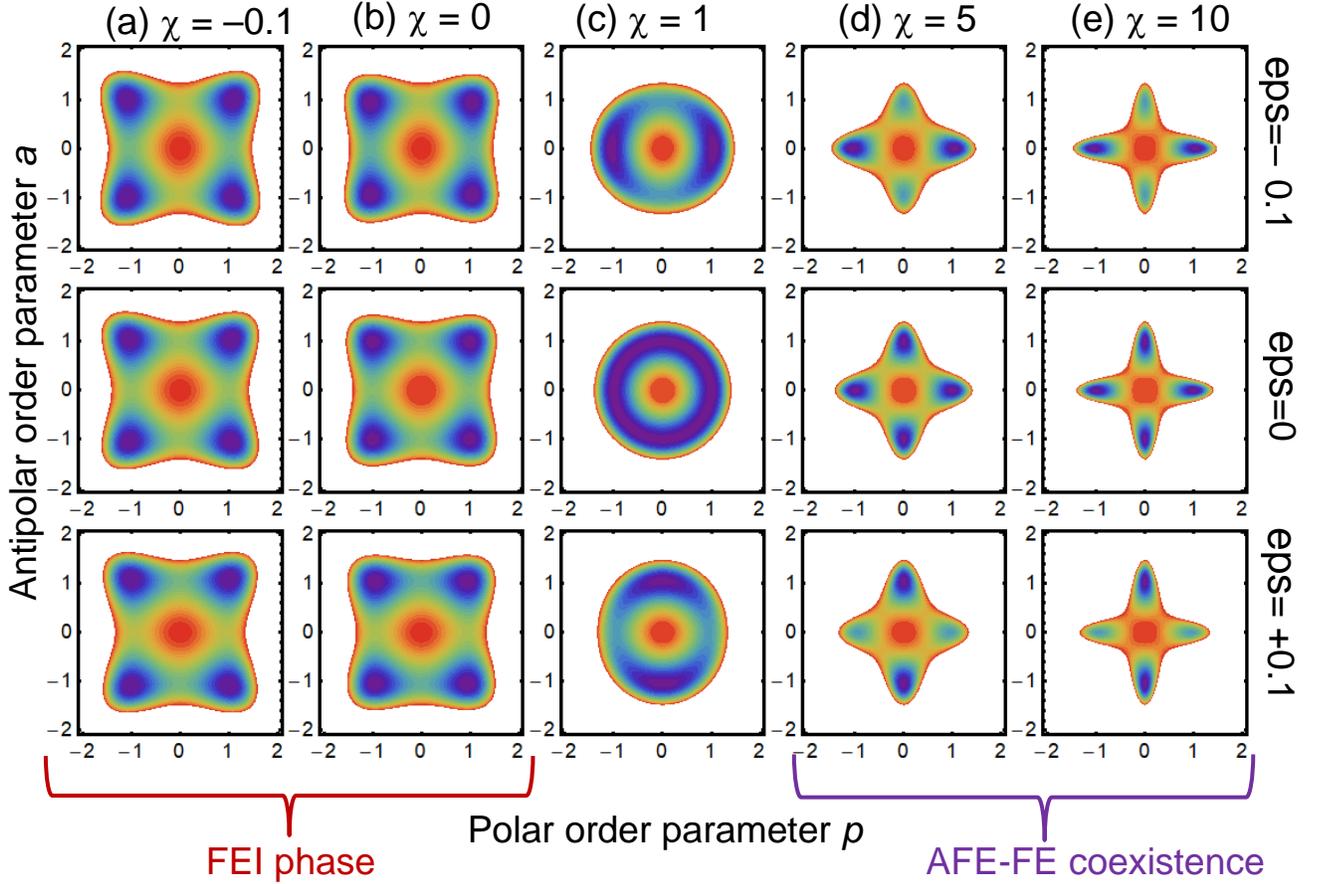

**FIGURE 1.** The free energy (6) as a function of polar ($p$) and antipolar ($a$) order parameters for different values of the AFE-FE coupling constant $\chi$: **(a)** $\chi = -0.1$, **(b)** $\chi = 0$, **(c)** $\chi = 1$, **(d)** $\chi = 5$ and $\chi = 10$ **(e)** and $|\epsilon| < 1$. Curie temperatures change parameter $\epsilon = -0.1$ for the top line, $\epsilon = 0$ for the middle line, $\epsilon = +0.1$ for the bottom line. Red color denotes zero energy, violet color is its minimal value in relative units.

The phase diagram containing the regions of the FE, AFE, FEI phases, as well as the stripped-colored region with coexisting AFE and FE states, is shown in **Fig. 2a** in dependence on the dimensionless doping factor $\epsilon$ and FE-AFE coupling constant $\chi$. The contour map of the dimensionless order parameters $a$ and $p$ in the coordinates $\epsilon$ and $\chi$ are shown in **Fig. 2b** and **2c**, respectively.

The FE R-phase, with $p \neq 0$ and $a = 0$, is stable at $\epsilon < 0$ and $\chi > \chi_{cr}^p(\epsilon)$ or $\chi < -1$. The AFE O-phases, with $a \neq 0$ and $p = 0$, are stable at $\epsilon > 0$, $\chi > \chi_{cr}^a(\epsilon)$ or $\chi < -1$. Both order parameters are nonzero in a mixed FEI phase, which is stable at $\epsilon < 0 \ \& \ -1 < \chi < \chi_{cr}^p(\epsilon)$, and $\epsilon > 0 \ \& \ -1 < \chi < \chi_{cr}^a(\epsilon)$.

Except for the calculated in this work mixed AFE-FE state located in the region $\chi > \frac{1+|\epsilon|}{1-|\epsilon|}$ and associated MP boundaries, **Figs. 2a** agrees with the phase diagram shown in Ref. [56].

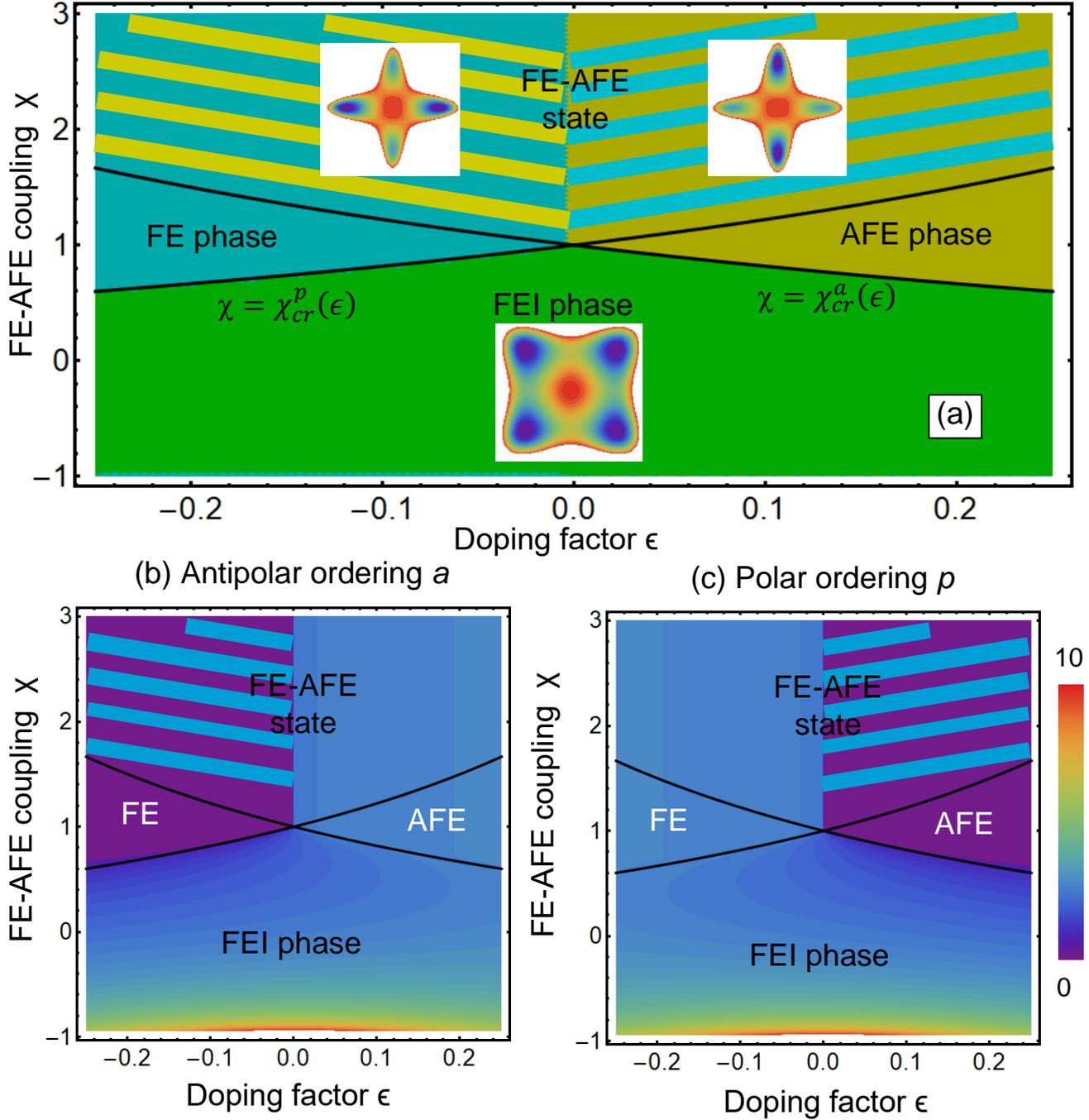

**FIGURE 2. (a)** The bulk phase diagram in coordinates "FE-AFE coupling strength $\chi$" and "doping factor $\epsilon$". The structure of the antiferrodistortive FE, AFE and FEI phases and coexisting FE and AFE states are listed in **Table I**.

Color maps of the dimensionless order parameters $a$ (**b**) and $p$ (**c**) in coordinates χ and ϵ. The color bar on the right represents the magnitude of the both order parameters.

Note that an LGD-type phenomenological model can describe the shape of the multi-well potential, but the microscopic origin of the order parameter(s), which form this potential, is not explained by the model [62]. Despite of the note, the main result of this section is the prediction of different 3D AFE-type long-range ordered phases in AFD multiferroics, such as $Bi_{1-y}R_yFeO_3$. The prediction will be compared with available experimental results in the next section.

## V. COMPARISON WITH EXPERIMENT

### A. X-ray Diffraction Experiments

Earlier we have shown that the SMPs in La doped BFO may correspond to the unit cell doubling and so may have different symmetries (see Refs. [55, 56] for details). It is natural to assume that the SMP phases in the multiferroic $Bi_{1-y}R_yFeO_3$ can have different symmetries corresponding to different structures of the polar-antipolar ordering. On atomic level, the SMPs structures are determined by different ordering of atomic displacements $\boldsymbol{U}^{(m)}$ [some of the cases are shown in **Table I**]. The prediction is compared with available XRD below.

As discussed in the introduction, the La-doped ceramics with the dopant concentration $0 < y < 0.45$ are characterized by the structural phase transition from the polar R-phase (space group R3c) to the antipolar O-phase *Pnam* (its supergroup *Pbam* can be considered to simplify calculations) (see **Fig. 3**) [63, 64]. The latter structure assumes an opposite alignment of the dipole moments oriented along the *a*-axis of the orthorhombic lattice (s.g. *Pbam*) having metric $\sqrt{2}a_p*2*\sqrt{2}a_p*2a_p$, where $a_p$ – is primitive perovskite cell parameter (see **Fig. 4**). The orthorhombic structure ascribed to the ceramic compounds $Bi_{1-y}La_yFeO_3$ in the concentration range $0.18 < y < 0.45$ is characterized by incommensurate modulation, which reaches its maximum value in the compounds with $y \sim 0.25$ [64]. It is known, that a decrease in the ionic radius of the dopant element in the systems $Bi_{1-y}R_yFeO_3$ leads to a reduction of the anti-polar displacement of the ions Bi(R), the concentration range ascribed to the anti-polar orthorhombic phase reduces thus leading to a destruction of the incommensurate modulation.

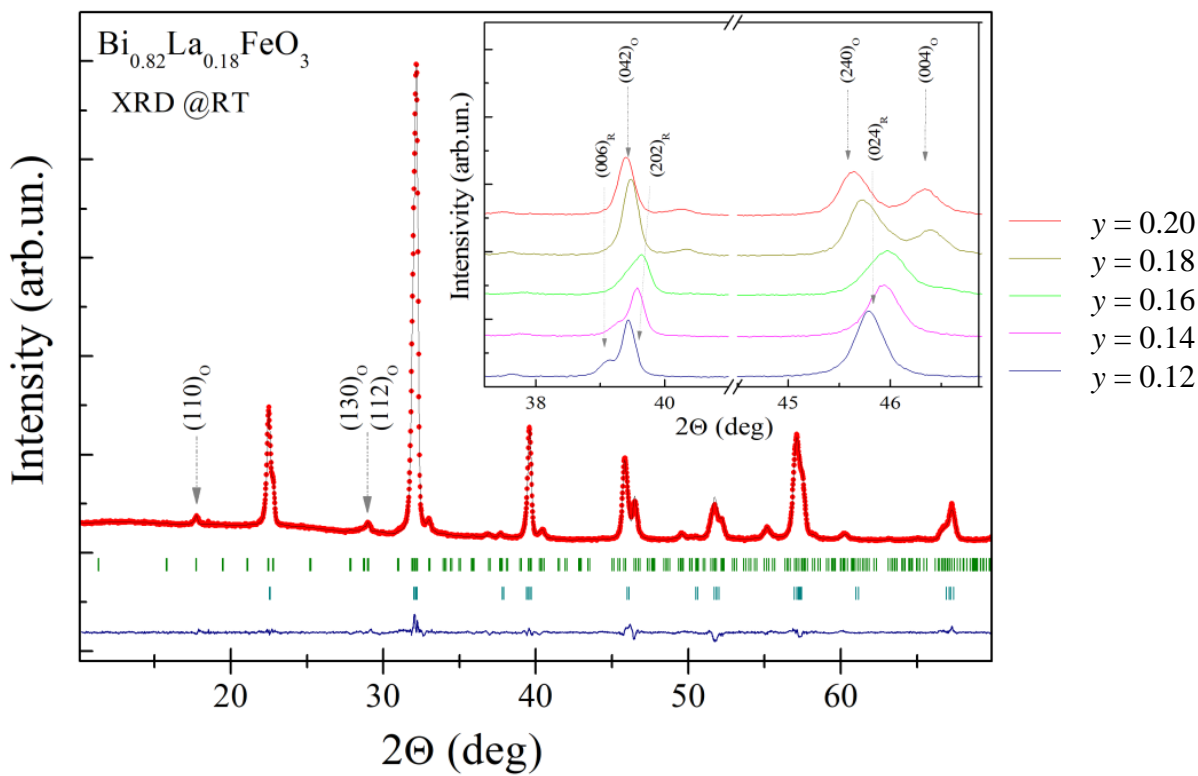

**FIGURE 3**. Room-temperature XRD pattern of compound BFO: La18% refined in the two-phase model. The upper row of the ticks denotes Bragg reflections ascribed to the orthorhombic phase, space group *Pbam*. The reflections specific for the R- and O- phases are indexed and shown on the inset for different dopant fraction *y*.

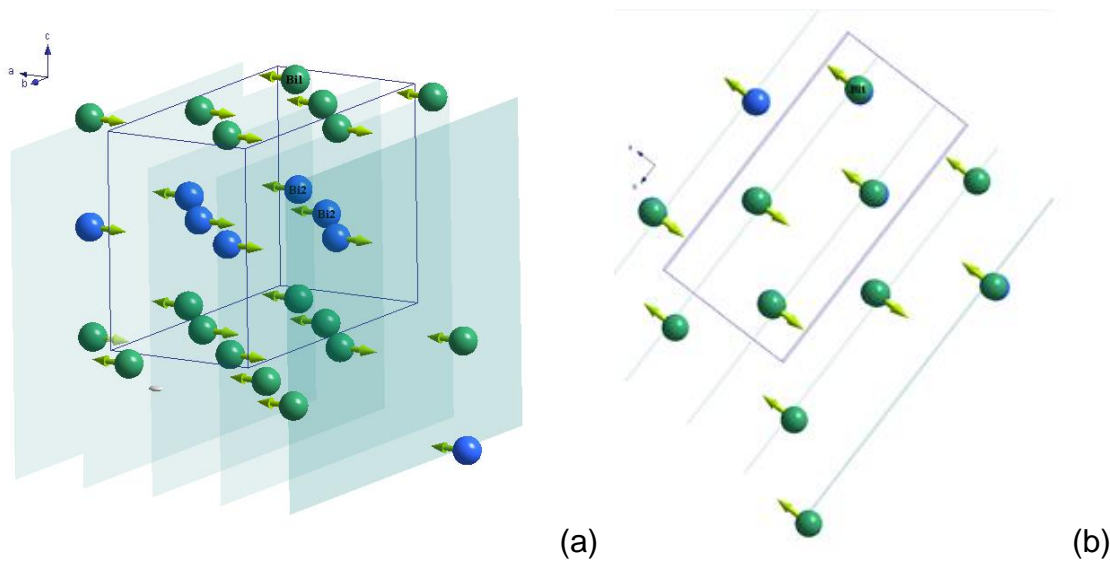

(a)                                                                 (b)

**FIGURE 4**. Visualization of the orthorhombic lattice (space group *Pbam*; metric $\sqrt{2}a_p*2*\sqrt{2}a_p*2a_p$), dipole moments associated with the Bi(R) ions are marked by arrows.

Sm-doped BiFeO₃ ceramics are characterized by a single-phase rhombohedral structure for the dopant content $y < 0.12$. The reflections attributed to the R-phase become negligible in the compounds with $y > 0.15$, thus the concentration range $0.12 \leq y \leq 0.15$ is attributed to the coexistence of the polar R- and the antipolar O-phases (see **Fig. 5**). Further increase in samarium content leads to the appearance of the new diffraction peaks ascribed to the nonpolar O-structure described by space group *Pnma*. The results of diffraction measurements obtained for the Sm-doped BFO ceramics could not reveal a modulation of the crystal structure, which can be caused by a limited resolution of the experimental set-up as well as an averaged character of the diffraction data, while local scale measurements presented below have clarified the structural data.

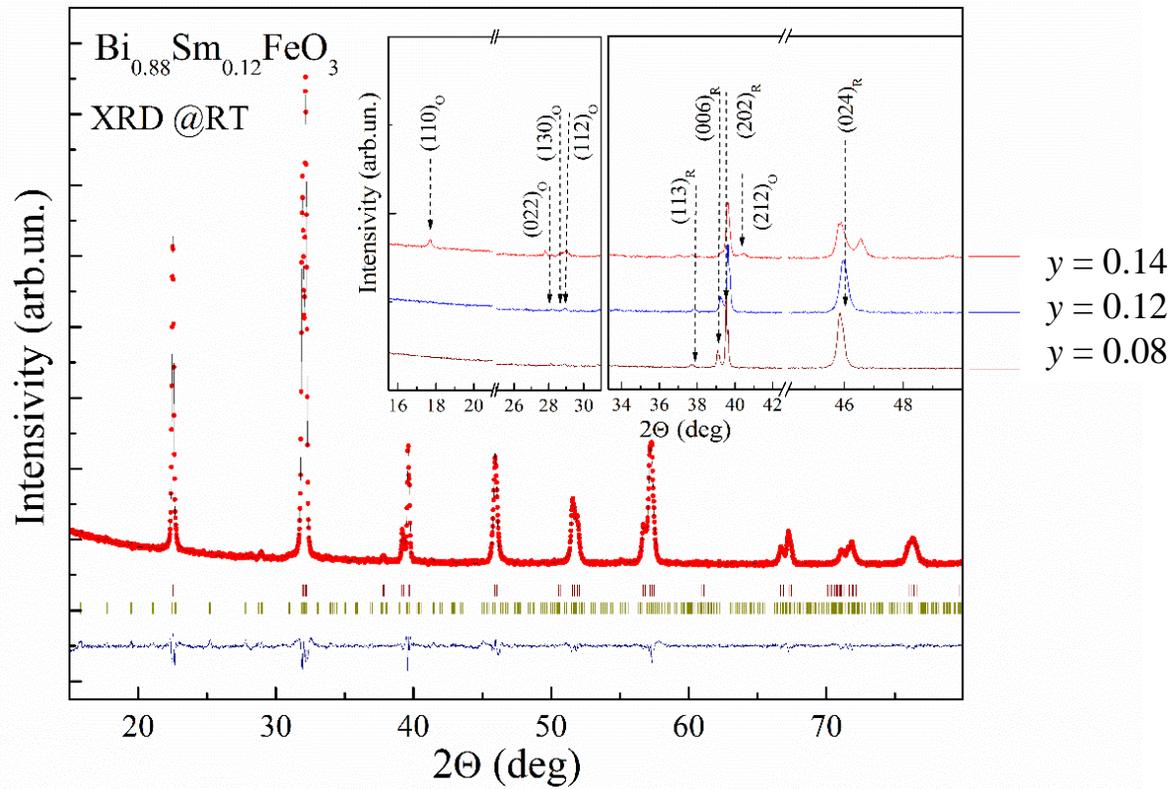

**FIGURE 5**. Room-temperature XRD pattern of compound Sm12%BFO refined in the two-phase model. The top row - Bragg reflections ascribed to the dominant R-phase, the second row – the O-phase, space group *Pbam*. The reflections specific for the R- and O- phases are indexed and shown on the inset for different dopant fraction *y*.

## B. Piezoresponse Force Microscopy Experiments

Typical PFM images of the three different compositions of Sm-doped BFO are represented on **Fig. 6**. The grain size was in a range of 0.5-2 μm, while the average size of the polar inclusions demonstrated clear trend of the polar phase size reduction (**Fig. 6**). Light and dark contrasts at the PFM images illustrate the existence of the randomly distributed polarization inside the grains of the samples, while dull uniform orange contrast is related to the regions without piezoresponse (**Figs 6**). The piezoelectric response was than analyzed and separated by the threshold of the noise level, around 1 pm. Piezoelectrically active (*R3c*) and inactive (antipolar *Pbam* or nonpolar *Pnma*) phases were extracted from the analysis of the PFM signal histograms [50] and plotted as separated images (**Fig. 6 d, h, l**). The average fraction of each phase was calculated by the estimation of the relative covered area (see **Table II,** the second row). The phase distribution clearly approves the phase transformation across the MPB with a decrease in the density of the piezoelectrically active regions and reduction of their average size. At 14 % of Sm doping the average size of the polar regions significantly reduces to 10-20 nm. This behavior is reliably coincident with the phenomenological description of the MPB, where the size of the piezoelectrically active regions decreases, and domain walls become significantly more mobile and contribute largely to the final electromechanical response.

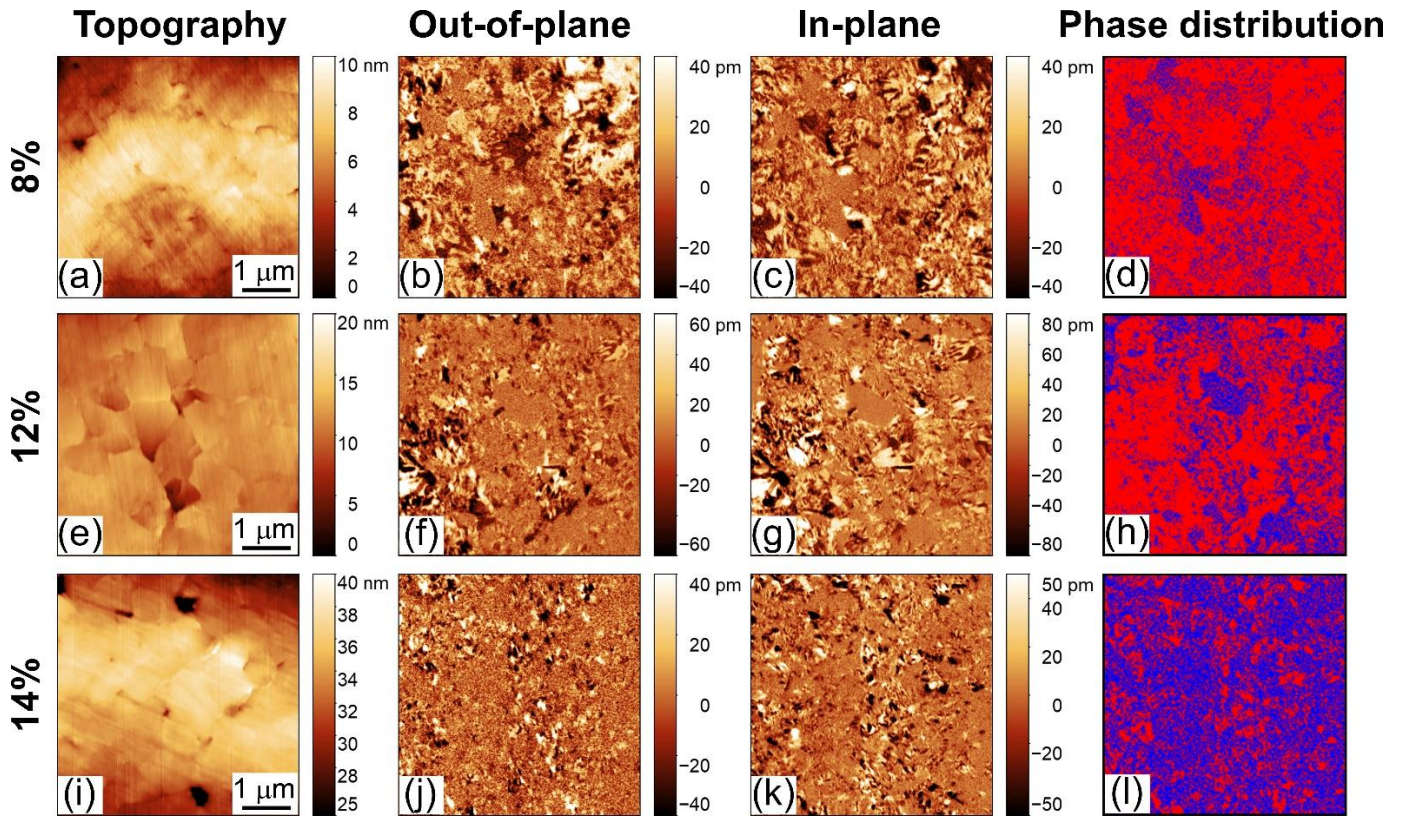

**FIGURE 6.** PFM and phase distribution images of BFO with different degree of Sm doping: **8% (the top row), 12% (the middle row), 14% (the bottom row).** The images **(a), (e), (i)** are topography; **(b), (f), (j)** are out-of-plane and **(c), (g), (k)** are in-plane PFM responses. Images **(d), (h),** (*l*) are phase distributions with piezoelectrically

active (red) and inactive (blue) phases, extracted by the comparative analysis of the out-of-plane and in-plane PFM images.

**Table II.** Phase content of the Sm doped BFO ceramics extracted from the PFM and XRD data

| Method | Phase ratio description | Sm doping (in %) of BFO | | |
|---|---|---|---|---|
| | | **8%** | **12%** | **14%** |
| PFM* | Polar/antipolar& nonpolar phase fraction (P/A) | 81/19 | 74/26 | 46/54 |
| XRD | Rhombohedral/orthorhombic phase ratio (R/O) | 100/0 | 90/10 | 20/80 |

*Accuracy of PFM measurements is 2%

The third row in **Table II** is the Rhombohedral/Orthorhombic (**R/O**) phase ratio, which is determined from the XRD data. Note that XRD data differentiates from the PFM data. It can be explained assuming the following argumentation. It is known that the sensitivity of the XRD phase analysis methods (i.e., the minimum amount of a phase in the case of coexistence of several structural phases) depends on the ratio of the absorption coefficients of the phases, size of crystallites, structural defects, mechanical stress etc. The accuracy of the XRD method for the BFO based compounds with mixed structural state is about $50 - 100$ nm, which is significantly lower than the PFM spatial resolution around 30 nm (size of the tip). The different sensitivities of these methods explain the differences in the R/O ratio estimated by the XRD and PFM methods. It should be also noted that small regions with O-AFE phase occur via the formation of nanosized nuclei and their growth in the vicinity of critical concentration $y_{cr}$ of the dopant ions. The randomly distributed nanoscale polar inclusions are "visible" by PFM, but indistinguishable by XRD method.

Typical PFM images of three different La compositions in BFO:La compounds are represented in **Fig. 7**. The piezoelectrically active areas represent regions of the rhombhedral *R3c* phase, while nonpiezoelectric areas are orthorhombic BFO phases: anti-polar *Pnam* or nonpolar *Pnma/Pbnm*. In 5% doped ceramics, most of the areas were piezoelectrically active except of the small areas inside holes, which can be attributed to the non-reliable PFM conditions. Domain walls in this composition were mostly straight. Increasing in the doping level leads to the change of the domain pattern: domain walls become more irregular and curved (**Figure 7, e-h**), while the polar phase area was reduced due to appearance of the small regions without piezoresponse. Both factors are the consequences of the substitution driven phase transition to the orthorhombic phase. In comparison to the Sm-doped BFO in the La-doped BFO compounds with x=18 % and 20 % the crystal structure is already predominantly orthorhombic (**Figure 7 i-p**). The regions of the polar active phase can be observed in the areas near the grains boundaries and topography defects which can be caused by slight chemical inhomogeneity formed through the crystallites volume as well as by mechanical stresses unavoidably induced in the regions near the defect location.

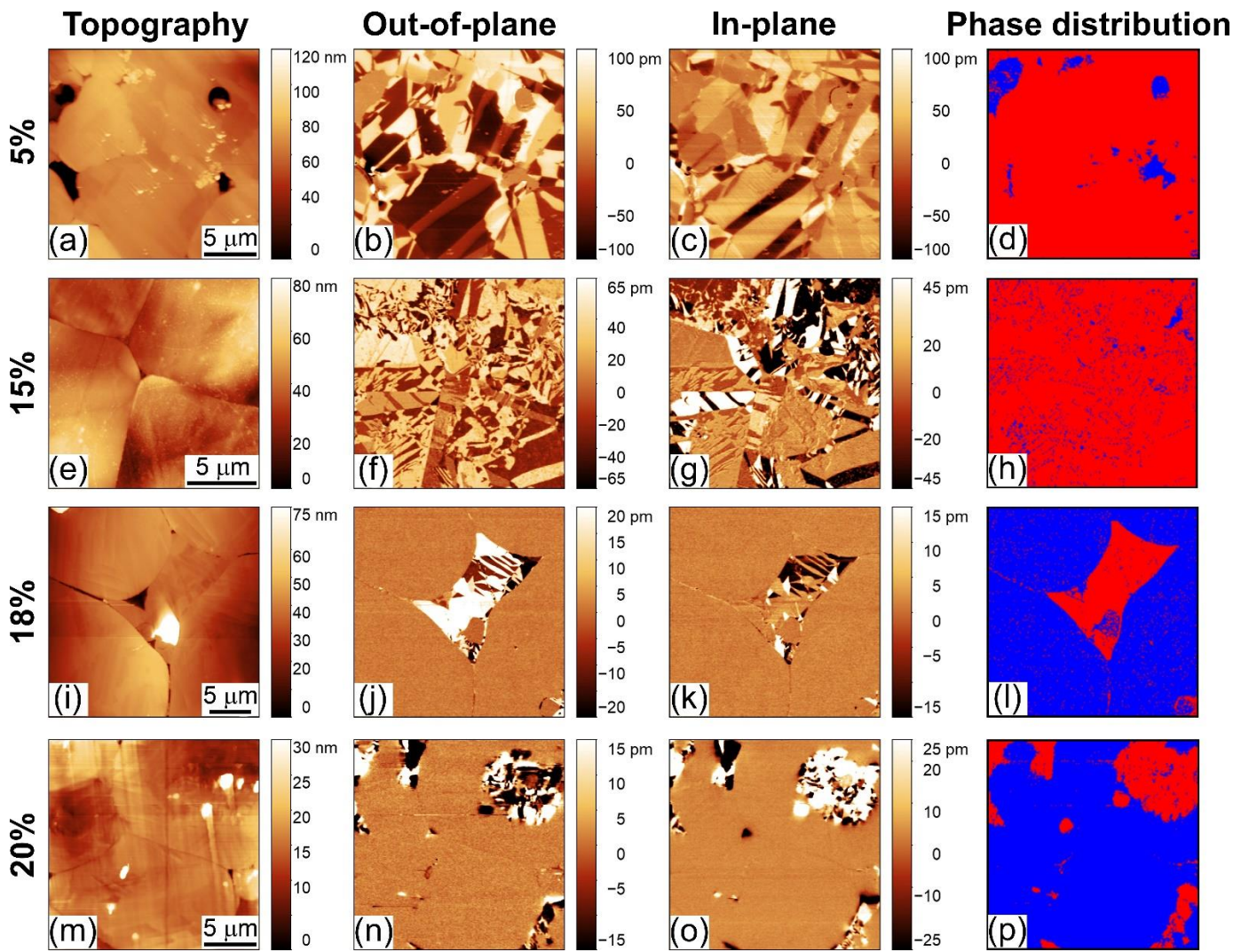

**FIGURE 7**. PFM and phase distribution images of BFO with different degree of La doping: **5% (the top row), 15% (the second row), 18% (the third row), and 20 % (the bottom row).** Images **(a), (e), (i), (m)** are topography; **(b), (f), (j), (n)** are out-of-plane and **(c), (g), (k), (o)** are in-plane PFM response. Images **(d), (h), (l), (p)** are the phase distribution with piezoelectrically active (red) and inactive (blue) phases, extracted by the comparative analysis of the out-of-plane and in-plane PFM images.

**Table III.** Phase content of the La doped BFO ceramics extracted from the PFM and XRD data

| Method | Phase ratio description | La doping (in %) of BFO ("±" means accuracy) | | | |
|---|---|---|---|---|---|
| | | **5%** | **15%** | **18%** | **22%** |
| PFM* | Polar/antipolar& nonpolar phase fraction (P/A) | 100±2/0±2 | 95±2/5±2 | 28±10/72±10 | 19±5/81±5 |
| XRD | Rhombohedral/orthorhombic phase ratio (R/O) | 100/0 | 100/0 | 20/80 | 10/90 |

*Accuracy of PFM measurements in dependence of La concentration is 2%, 2%, 10% and 5%, respectively

## C. Fitting of PFM and XRD Results

The ratio $\mu$ of orthorhombic AFE to rhombohedral FE phase content in the BFO:Sm and BFO:La ceramics is shown in **Fig. 8a** and **8b**, respectively. Points represent experimental data from **Table II** and **III**, respectively, while solid curves are a theoretical fit with an exponential-type functions:

$$\mu(y) = \begin{cases} C_1 \left[ \exp\left( \frac{y - y_{cr}}{\Delta_1} \right) - 1 \right] + C_2 \left[ 1 - \exp\left( -\frac{y - y_{cr}}{\Delta_2} \right) \right], & y_{cr} \leq y \leq 1, \\ 0, & y < y_{cr}. \end{cases} \quad (7)$$

Here $y$ is the chemical content of Sm or La content in %, $y_{cr}$ is a critical concentration of O-AFE phase nucleation, amplitudes $C_{1,2}$ and characteristic concentrations $\Delta_{1,2}$ are fitting parameters listed in **Table IV**. Note that the exponential functions are typical functions for the description of nucleation process of the emerging phase critical nuclei.

**Table IV.** Fitting parameters in Eq.(7) for the Sm and La doped BFO ceramics

| Fitting parameter (in %) | $y_{cr}$ (%) | $C_1$ (%) | $\Delta_1$ (%) | $C_2$ (%) | $\Delta_2$ (%) |
|---|---|---|---|---|---|
| BFO:Sm ceramics (from PFM data) | 7 | 0.0294 | 1 | 21.7827 | 0.5 |
| BFO:Sm ceramics (from XRD data) | 7 | 0.0729 | 1 | 0.0000 | N/A |
| BFO:La ceramics (from PFM data) | 15 | 0.0080 | 1 | 72.09 | 0.5 |
| BFO:La ceramics (from XRD data) | 15 | 0.0091 | 1 | 80.0246 | 0.5 |

Note that the difference between the blue and cyan curves in **Fig. 8b** are relatively small, at least much smaller than the difference between corresponding red and orange curves in **Fig. 8a**. This circumstance reflects the difference between PFM and XRD measurements of AFE/FE phase ratio in BFO:Sm, while PFM and XRD measurements for BFO:La give relatively close values. In a definite sense both fittings of PFM data and one fitting of XRD data for BFO:La give qualitatively and semi-quantitatively close results – two exponents with the same $\Delta_1$ and $\Delta_2$ values. Only XRD data for BFO:Sm is fitted by one exponent, and the form of the orange fitting curve is different from the red, blue and cyan curves. The discrepancy may originate from the fact that XRD is insensitive to the small ($< 10$ nm) inclusions of emerging phase (in the considered case AFE (O) phase) which concentration is smaller than minimal amount necessary to provide notable reflection on the diffraction pattern. So, XRD underestimates the concentration of O-phase onset. At the same time PFM can resolve such small inclusions, while their apparent size can be significantly bigger than the real size due to the tip resolution. In this sense PFM with a tip size about 30 nm overestimates the AFE/FE ratio at the AFE phase onset.

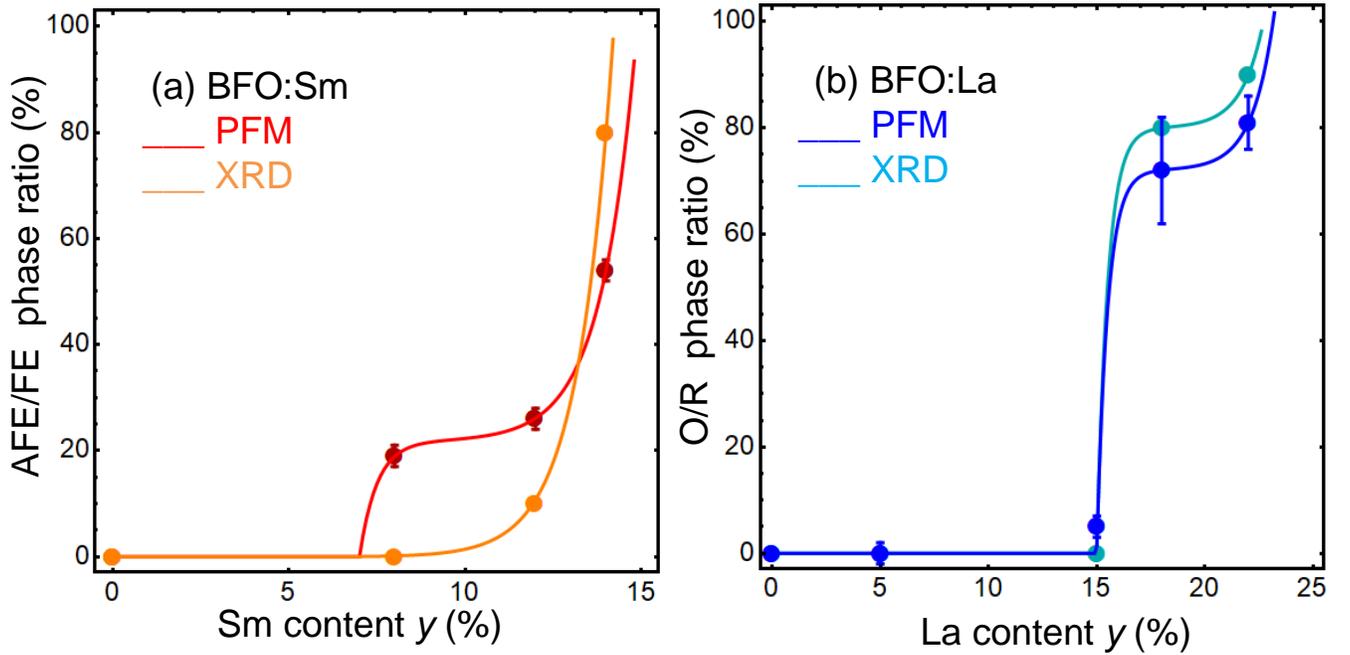

**FIGURE 8.** **(a)** The ratio of orthorhombic AFE (O) to rhombohedral FE (R) phase content in the BFO:Sm ceramics. Red and orange symbols are experimental data from **Table II**, red and orange solid curve is a theoretical fit by Eq.(7) of PFM and XRD results, respectively. **(b)** The ratio of AFE-O to FE-R phase content in the BFO:La ceramics. Blue and cyan symbols are experimental data from **Table III**, blue and cyan solid curves are a theoretical fit by Eq.(7) of PFM and XRD results, respectively.

From the fitting results we must conclude that we cannot limit the functional form in Eq.(3a) by the linear terms to obtain good fitting. This result is intuitively clear, because we need to describe the complete phase diagram, where the fraction of AFE-O and FE-R phases changes from 0 to 100%.

## D. Probability Analysis

The unconditional probabilities $W_{AFE}$, $W_{FE}$ and $W_{FEI}$ of the O-AFE, R-FE and FEI phases realization in a localized spatial region of a BFO:Sm and BFO:La ceramics with a local variation of chemical composition, grain boundaries, etc. are expressed via the energy levels $G_{AFE}$, $G_{FE}$, $G_{FEI}$, and canonic statistic sum $Z$ in a conventional way:

$$W(AFE) = \frac{1}{Z} exp \left[ -\frac{G_{AFE}}{k_B T} \right], \quad W(FE) = \frac{1}{Z} exp \left[ -\frac{G_{FE}}{k_B T} \right], \quad W(FEI) = \frac{1}{Z} exp \left[ -\frac{G_{FEI}}{k_B T} \right], \tag{8a}$$

$$Z = exp \left[ -\frac{G_{AFE}}{k_B T} \right] + exp \left[ -\frac{G_{FE}}{k_B T} \right] + exp \left[ -\frac{G_{FEI}}{k_B T} \right]. \tag{8b}$$

Here the free energies $G_{FE} = -\frac{G_0}{4} (1-\epsilon)^2$, $G_{AFE} = -\frac{G_0}{4} (1+\epsilon)^2$ and $G_{FEI} = -G_0 \frac{1+\epsilon^2 - \chi(1-\epsilon^2)}{2(1-\chi^2)}$ per **Table I**, and the energy value $G_0 = \alpha P_S^2$ is referred to a pure bulk BFO. The conditional probabilities of the observable AFE state realization are given by other expressions, which depend on a definite hypothesis.

The **first hypothesis** is that the ceramics is in a coexisting AFE-FE state, but not in a FEI or other pure phase. In this case the conditional probabilities of AFE and FE local states are:

$$W(AFE|AFE \cup FE) = \frac{exp\left[-\frac{G_{AFE}}{k_B T}\right]}{exp\left[-\frac{G_{AFE}}{k_B T}\right] + exp\left[-\frac{G_{FE}}{k_B T}\right]}, \quad W(FE|AFE \cup FE) = \frac{exp\left[-\frac{G_{FE}}{k_B T}\right]}{exp\left[-\frac{G_{AFE}}{k_B T}\right] + exp\left[-\frac{G_{FE}}{k_B T}\right]}. \tag{9a}$$

The ratio $\mu$, which is measured experimentally, can regarded equal to the ratio of the AFE and FE states conditional probabilities:

$$\mu = exp\left[\frac{G_{FE} - G_{AFE}}{k_B T}\right] = exp\left[\frac{G_0}{k_B T}\epsilon\right], \tag{9b}$$

Hence the y-dependence of the doping parameter $\epsilon$ can be estimated from Eqs.(7) and (9b) and $y > y_{cr}$ as:

$$\epsilon(y) = \frac{k_B T}{G_0} ln[\mu] \approx \frac{k_B T}{G_0} ln \left[ C_1 \left[ exp\left(\frac{y - y_{cr}}{\Delta_1}\right) - 1 \right] + C_2 \left[ 1 - exp\left(-\frac{y - y_{cr}}{\Delta_2}\right) \right] \right]. \tag{9c}$$

The dependence of $\epsilon(y)$ deconvoluted from the PFM and XRD results for $\mu(y)$ are shown in **Fig. 9a** and **9b** for BFO:Sm and BFO:La, respectively. Note that in this case the doping parameter $\epsilon$ is $\chi$ -independent and can be unambiguously determined from the experiments.

**Fig. 9c** and **9d** are reconstructed phase diagrams of BFO:Sm and BFO:La in coordinates "doping content y – FE-AFE coupling constant $\chi$". It is seen that both diagrams contain a FE R-phase at $y < y_{cr}$, which borders with coexisting FE-AFE state at $\chi > 1$ (the first vertical MP boundary), and with FEI phase at $\chi < 1$ (the second vertical MP boundary). A very short vertical boundary between FE and AFE phase is located at the vicinity $\chi = 1$. The AFE O-phase has extended curved boundaries with FE-AFE and FEI phases.

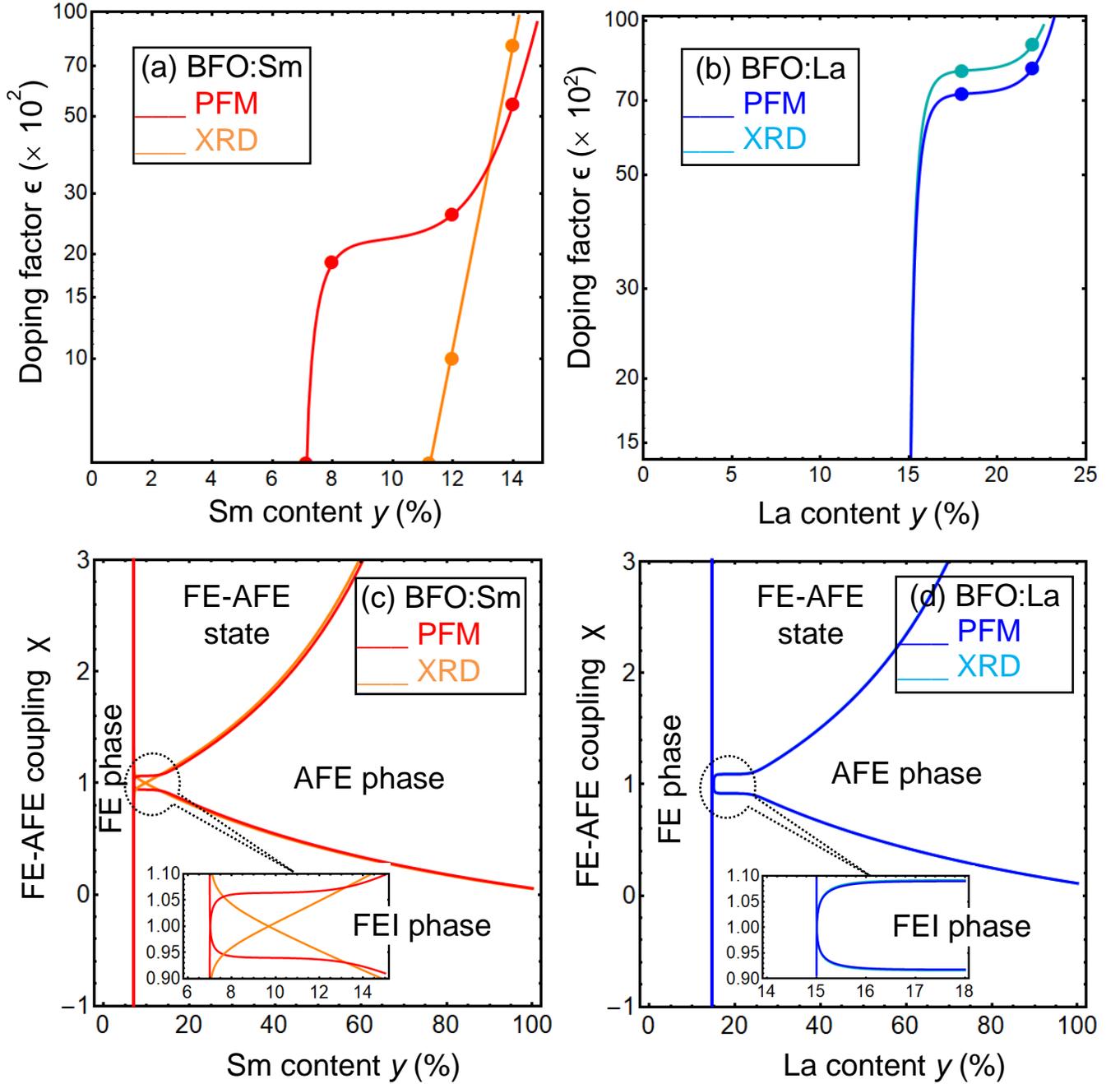

**FIGURE 9**. **(a)** The doping parameter $\epsilon$ of the BFO:Sm ceramics. Red and orange symbols are the points deconvoluted from the experimental data in **Table II**, red and orange solid curves are a theoretical fit by Eq.(9c) for PFM and XRD results, respectively. **(b)** The doping parameter $\epsilon$ of the BFO:La ceramics. Blue and cyan symbols are deconvoluted from the experimental data in **Table III**, blue and cyan solid curves are a theoretical fit by Eq.(9c) for PFM and XRD results, respectively. The scaling factor $\frac{k_B T}{G_0} = 0.01$. Plots **(c)** and **(d)** are reconstructed from XRD and PFM data phase diagrams of BFO:Sm and BFO:La in coordinates "doping content y – FE-AFE coupling constant $\chi$". Red and orange,

blue and cyan solid curves are plotted for the parameters in **Table IV**. Note that blue and cyan solid curves in the part (d) coincide. Insets in the parts (c,d) are zoomed parts of the corresponding phase diagrams.

Note that red and orange, blue and cyan solid curves in **Fig. 9c** and **9d** are plotted for the parameters in **Table IV** corresponding to PFM and XRD results, respectively. Note that the difference between the red and orange curves in **Fig. 9c** are very small, while the difference between corresponding PFM and XRD data fitting is quite visible (see red and orange curves in **Fig. 8a**). This circumstance reflects the stability of the proposed fitting procedure for BFO:Sm. The blue and cyan solid curves in **Fig. 9d** coincide, because the corresponding PFM and XRD data fitting for BFO:La are relatively close (see blue and cyan curves in **Fig. 8a**). Hence the complementary XRD and PFM methods, and theoretical reconstruction of the doping factor allow us to recover the more complete picture of phase coexistence in BFO:R.

The **second hypothesis** is that the ceramics can be near the boundary between AFE and FEI phases, and not in a coexisting AFE-FE state. In the much rarer case, the conditional probabilities of AFE and FEI local states are given in **Appendix B.** It appeared that the approximate analytical y-dependence of the doping factor $\epsilon$ can be derived only at very small $|\epsilon| \ll 1$ and $y > y_{cr}$, and it is $\chi$-dependent:

$$\epsilon(y) = 2\frac{k_B T}{G_0}ln[\mu(y)] - \frac{\chi-1}{2(\chi+1)}. \tag{10}$$

Thus, in this case the $\chi$-dependent doping parameter cannot be unambiguously determined from PFM and/or XRD experiments. Therefore, the model is not used by us.

The causal [51] or/and Bayesian [65] analyses based on Eqs (9) and (10) can be in order to consider both (or more) hypothesis together, but the applicability of these approaches goes well beyond the used thermodynamic approach.

## VI. CONCLUSION

Using LGD phenomenology, semi-microscopic FSM model, XRD and PFM experiments, this work predicts the emergence and explores the behavior of AFE, FEI and FE long-range ordering and morphotropic AFE-FE phase boundaries in AFD rare-earth doped multiferroics, such as $Bi_{1-y}R_yFeO_3$, where $R$ is Sm or La atoms.

In accordance with the XRD data the BFO compounds doped with La ions in the concentration range $0.16 < y < 0.18$ are characterized by a coexistence of the polar R- and the antipolar O- phases followed by the single phase antipolar orthorhombic state, wherein the orthorhombic phase possesses an incommensurate modulation. In the Sm-doped compounds the R/O phase coexistence corresponds to the concentration range $0.12 \leq y \leq 0.15$, wherein the antipolar O-phase does not have any evidences of the modulation. The concentration ranges of the MPB regions estimated by the diffraction data notably differ from those determined by PFM measurements. Such a behavior is typical for the MP boundary and does not significantly depend on the sintering technique. More extended phase coexistence region observed by PFM method for the Sm-doped compounds as compared to that determined by the diffraction measurements is caused by different measurement limits ascribed to these techniques. The compounds doped with Sm ions are characterized by polar active and non-polar inclusions associated respectively with the rhombohedral and the orthorhombic phases have nanoscale size and thus cannot be reliably detected by the diffraction measurements while can be observed by PFM method. In the case of La-doped compounds the MP boundary regions estimated by the diffraction data and the PFM results are nearly equal as the average size of the polar active and non-polar inclusions is more than 100 nm and these inclusions can be reliably detected by both experimental methods.

We concluded that the observed nontrivial phase coexistence originates from the dopant interaction with the soft mode dipoles (i.e. with the polar order), which provoke the AFE-FE type interaction between neighboring polar sublattices with the doping factor increase. To quantify the conclusion, we minimize LGD-FSM free energy with two master parameters, which are the doping factor $\epsilon$ and the coupling strength $\chi$ between four neighboring A-sites in a $BiFeO_3$ structure. The dependences of $\epsilon(y)$ were deconvoluted from the PFM and XRD results for the AFE/FE phase ratio in the BFO:Sm and BFO:La, respectively. Using the deconvoluted dependence $\epsilon(y)$, the phase diagrams of BFO:Sm and BFO:La were reconstructed in coordinates "doping content y – FE-AFE coupling constant $\chi$". The diagrams contain a FE R-phase at $y < y_{cr}$, which borders with coexisting FE and AFE states at $\chi > 1$ (the first MP boundary), and with FEI phase at $\chi < 1$ (the second MP boundary). A very short vertical boundary between FE and AFE phase is located at the vicinity of $\chi = 1$. The AFE O-phase has extended curved boundaries with FE-AFE and FEI phases. Thus, the possibility of the LGD-FSM approach to describe qualitatively the experimental results obtained by XRD and PFM measurements for $Bi_{1-y}La_yFeO_3$ and $Bi_{1-y}Sm_yFeO_3$ ceramics is demonstrated.

The complementary XRD and PFM methods, and theoretical reconstruction of the doping factor allow us to recover the more complete picture of phase coexistence in BFO:R. The proposed combined theoretical and experimental approach, thereby, provides a promising method to the inspect an origin of FE and AFE phase MPB of the in rare-earth doped multiferroics.

**Acknowledgements.** Authors acknowledge Dr. Bobby Sumpter (ORNL) for useful suggestions. This material is based upon work (S.V.K.) supported by the U.S. Department of Energy, Office of Science, Office of Basic Energy Sciences, and performed at the Center for Nanophase Materials Sciences, a US Department of Energy Office of Science User Facility. A portion of FEM was conducted at the Center for Nanophase Materials Sciences, which is a DOE Office of Science User Facility (CNMS Proposal ID: 257). A.N.M. work is supported by the National Academy of Sciences of Ukraine (the Target Program of Basic Research of the National Academy of Sciences of Ukraine "Prospective basic research and innovative development of nanomaterials and nanotechnologies for 2020 - 2024", Project № 1/20-H, state registration number: 0120U102306). A.N.M., D.V.K., A.D.Y., O.M.F., T.S., V.V.S. and A.L.K. received funding from the European Union's Horizon 2020 research and innovation programme under the Marie Skłodowska-Curie grant agreement No 778070. A.N.M. acknowledges the National Research Foundation of Ukraine. M.V.S. acknowledges financial support from the Ministry of Science and Higher Education of the Russian Federation within the framework of state support for the creation and development of World-Class Research Centers "Digital biodesign and personalized healthcare" №075-15-2020-926. Part of the work (A.L.K.) was supported by the Ministry of Education and Science of the Russian Federation in the framework of the Increase Competitiveness Program of NUST «MISiS» (No. K2-2019-015). V.V.S. and A.L.K. were additionally supported by RFBR and BRFBR, project numbers 20-58-0061 and T20R-359, respectively. Part of this work (A.L.K.) was developed within the scope of the project CICECO-Aveiro Institute of Materials, refs. UIDB/50011/2020 and UIDP/50011/2020, financed by national funds through the FCT/MCTES.

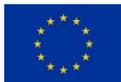

**Authors' contribution.** A.N.M. generated the research idea, proposed the theoretical model and obtained analytical results. D.V.K. and M.V.S. conducted XRD measurements and interpreted obtained results. D.O.A. and A.A. performed PFM measurements and interpreted obtained results. E.A.E. and A.N.M. fitted XRD and PFM data. A.P. and A.K. selected technological regimes for making samples, prepared and characterized the samples. A.D.Y., O.M.F., T.V.S. and V.V.S. selected technological regimes for making samples. A.N.M., D.V.K., D.O.A., and M.V.S. wrote the manuscript draft. M.D.G., S.V.K. and A.L.K. worked on the results discussion and manuscript improvement.

# Supplementary Materials

## APPENDIX A. Conditions of the Phase Stability

The free energy (10) as a function of polar and antipolar order parameters, $p$ and $a$, is:

$$g_{LGD} = -(1-\epsilon)\frac{p^2}{2} - (1+\epsilon)\frac{a^2}{2} + \frac{p^4+a^4}{4} + \frac{\chi}{2}p^2a^2 + \gamma\frac{p^6+a^6}{6} + \frac{g}{2}\left[\left(\frac{dp}{dx}\right)^2 + \left(\frac{da}{dx}\right)^2\right], \quad \text{(A.1)}$$

Note that the conditions $\gamma \geq 0$ and $g > 0$ should be fulfilled for the stability of the functional (A.1). At that we must put $\chi > -1$ for the case $\gamma = 0$. Also, we will regard that $|\epsilon| < 1$

The properties of bulk homogeneous system with $dp/dx = 0$ and $da/dx = 0$ are considered below. The equations of state are

$$-(1-\epsilon)p + p^3 + \chi\, p\, a^2 + \gamma p^5 = 0, \quad \text{(A.2a)}$$

$$-(1+\epsilon)a + a^3 + \chi\, p^2 a + \gamma a^5 = 0, \quad \text{(A.2b)}$$

The solution of (A.2a)-(A.2b) determine the phase of the system. Only the stable solutions are physically permissible, hence the matrix of the second derivatives of the free energy (A.1) with respect to order parameters $a$ and $p$ should be positively defined:

$$\left\|\frac{\partial^2 g_L}{\partial p\, \partial a}\right\| = \begin{pmatrix} -(1-\epsilon) + 3p^2 + 5\gamma p^4 + \chi a^2 & 2\chi\, p\, a \\ 2\chi\, p\, a & -(1+\epsilon) + 3a^2 + 5\gamma a^4 + \chi p^2 \end{pmatrix} \quad \text{(A.3a)}$$

For the matrix to be positively defined the following conditions must be valid:

$$-(1-\epsilon) + 3p^2 + 5\gamma p^4 + \chi a^2 > 0, \quad -(1+\epsilon) + 3a^2 + 5\gamma a^4 + \chi p^2 > 0, \quad \text{(A.3b)}$$

$$[3a^2 + 5\gamma a^4 + \chi p^2 - (1+\epsilon)][3p^2 + 5\gamma p^4 + \chi a^2 - (1-\epsilon)] - 4\chi^2 p^2 a^2 > 0. \quad \text{(A.3c)}$$

Four spatially homogeneous phases could exist for the system with free energy (A.1), namely:

1) Paraelectric **PE** phase with the order parameters and energy density

$$p = 0, \quad a = 0 \quad \text{(A.4a)}$$

$$g_L = 0 \quad \text{(A.4b)}$$

Stability matrix (A.3) for this case is

$$\left\|\frac{\partial^2 g_L}{\partial p\, \partial a}\right\| = \begin{pmatrix} -(1-\epsilon) & 0 \\ 0 & -(1+\epsilon) \end{pmatrix} \quad \text{(A.4c)}$$

Hence PE phase is unstable at $|\epsilon| < 1$.

2) For $\gamma = 0$ and $\chi > -1$ the polar **FE** phase is characterized by the order parameters:

$$p = \pm\sqrt{1-\epsilon}, \quad a = 0, \quad \text{for} \quad \gamma = 0. \quad \text{(A.5a)}$$

$$p = \pm\sqrt{\frac{-1+\sqrt{1+4\gamma(1-\epsilon)}}{2\gamma}}, \quad a = 0, \quad \text{for} \quad \gamma > 0 \quad \text{(A.5b)}$$

The energy density and stability matrix (A.3) for the case $\gamma = 0$ and $\chi > -1$ is

$$g_L = -\frac{(1-\epsilon)^2}{4} \tag{A.5c}$$

$$\left\|\frac{\partial^2 g_L}{\partial p\, \partial a}\right\| = \begin{pmatrix} 2(1-\epsilon) & 0 \\ 0 & -(1+\epsilon)+\chi(1-\epsilon) \end{pmatrix} \tag{A.5d}$$

Hence, FE is stable at

$$\chi > \chi_{cr}^p(\epsilon) \overset{\text{def}}{=} \frac{1+\epsilon}{1-\epsilon} \qquad \text{and} \qquad \epsilon < 1 \tag{A.5e}$$

3) Antipolar **AFE** phase with the order parameters and energy density:

$$p = 0, \; a = \pm\sqrt{1+\epsilon} \quad \text{for} \quad \gamma = 0. \tag{A.6a}$$

$$p = 0, \quad a = \pm\sqrt{\frac{-1+\sqrt{1+4\gamma(1+\epsilon)}}{2\gamma}}, \quad \text{for} \quad \gamma > 0 \tag{A.6b}$$

The energy density and stability matrix (A.3) for the case $\gamma = 0$ and $\chi > -1$ is

$$g_L = -\frac{(1+\epsilon)^2}{4}, \tag{A.6c}$$

$$\left\|\frac{\partial^2 g_L}{\partial p\, \partial a}\right\| = \begin{pmatrix} -(1-\epsilon)+\chi(1+\epsilon) & 0 \\ 0 & 2(1+\epsilon) \end{pmatrix} \tag{A.6c}$$

Hence, AFE phase is stable at

$$\chi > \chi_{cr}^a(\epsilon) \overset{\text{def}}{=} \frac{1-\epsilon}{1+\epsilon} \qquad \text{and} \qquad \epsilon > -1 \tag{A.6d}$$

4) For $\gamma = 0$ and $\chi > -1$ the mixed ferrielectric **FEI** phase with the order parameters and energy density:

$$p = \pm\sqrt{\frac{1-\epsilon-\chi(1+\epsilon)}{1-\chi^2}}, \; a = \pm\sqrt{\frac{1+\epsilon-\chi(1-\epsilon)}{1-\chi^2}}, \tag{A.7a}$$

$$g_L = -\frac{1+\epsilon^2-\chi(1-\epsilon^2)}{2(1-\chi^2)}. \tag{A.7b}$$

Allowing for equations of state (A.2) at $p^2 > 0, a^2 > 0$, the stability matrix (A.3a) for this case is

$$\left\|\frac{\partial^2 g_L}{\partial p\, \partial a}\right\| = \begin{pmatrix} 2p^2 & 2\chi\, p\, a \\ 2\chi\, p\, a & 2a^2 \end{pmatrix} \tag{A.7c}$$

One could see from (A.7c) that FEI-phase is stable at $p^2 > 0, a^2 > 0$ and $\chi^2 < 1$, which gives the following conditions:

$$|\epsilon| < \frac{1-\chi}{1+\chi} \; \text{ and } \; \chi^2 < 1, \text{ which is equivalent to } \quad \chi < \frac{1-|\epsilon|}{1+|\epsilon|} \text{ and } |\epsilon| < 1. \tag{A.7d}$$

5) FE-AFE equilibrium is achieved under the condition $\epsilon = 0$ at $\chi > 1$; FEI-AFE equilibrium is achieved under the condition $\chi = \chi_{cr}^a(\epsilon)$ at $\chi^2 < 1$, FEI-FE equilibrium is achieved under the condition $\chi = \chi_{cr}^p(\epsilon)$ at $\chi^2 < 1$.

The coexistence of FE and AFE states is possible under the condition:

$$\chi > \frac{1+|\epsilon|}{1-|\epsilon|} \text{ and } |\epsilon| < 1, \text{ which is equivalent to } \quad |\epsilon| < -\frac{1-\chi}{1+\chi} \text{ and } \chi > 1. \tag{A.8}$$

### APPENDIX B. Probability Analysis

The unconditional probabilities $W_{AFE}$, $W_{FE}$ and $W_{FEI}$ of the O-AFE, R-FE and FEI phases realization in a localized spatial region of a doped BFO ceramics with a local variation of chemical composition, grain boundaries, etc. are expressed via the energy levels $G_{AFE}$, $G_{FE}$, $G_{FEI}$, and canonic statistic sum $Z$ in a conventional way:

$$W(AFE) = \frac{1}{Z} exp\left[-\frac{G_{AFE}}{k_B T}\right], \quad W(FE) = \frac{1}{Z} exp\left[-\frac{G_{FE}}{k_B T}\right], \quad W(FEI) = \frac{1}{Z} exp\left[-\frac{G_{FEI}}{k_B T}\right], \quad \text{(B.1a)}$$

$$Z = exp\left[-\frac{G_{AFE}}{k_B T}\right] + exp\left[-\frac{G_{FE}}{k_B T}\right] + exp\left[-\frac{G_{FEI}}{k_B T}\right]. \quad \text{(B.1b)}$$

Where the free energies $G_{FE} = -\frac{G_0}{4}(1-\epsilon)^2$, $G_{AFE} = -\frac{G_0}{4}(1+\epsilon)^2$ and $G_{FEI} = -G_0 \frac{1+\epsilon^2 - \chi(1-\epsilon^2)}{2(1-\chi^2)}$ per **Table I**, and the energy value $G_0 = \alpha P_S^2$ is referred to a pure bulk BFO.

The conditional probabilities of the observable AFE state realization are given by other expressions, which depend on a definite hypothesis.

The **first hypothesis** is that the ceramics is in a coexisting AFE-FE state, but not in a FEI and phase. In this case the conditional probabilities of AFE and FE local states are:

$$W(AFE|AFE \cup FE) = \frac{exp\left[-\frac{G_{AFE}}{k_B T}\right]}{exp\left[-\frac{G_{AFE}}{k_B T}\right] + exp\left[-\frac{G_{FE}}{k_B T}\right]}, \quad W(FE|AFE \cup FE) = \frac{exp\left[-\frac{G_{FE}}{k_B T}\right]}{exp\left[-\frac{G_{AFE}}{k_B T}\right] + exp\left[-\frac{G_{FE}}{k_B T}\right]}. \quad \text{(B.2a)}$$

The ratio $\mu$, which is measured experimentally, can regarded equal to the ratio of the AFE and FE states probabilities calculated theoretically:

$$\mu = \frac{w(AFE|AFE \cup FE)}{w(FE|AFE \cup FE)} = exp\left[\frac{G_{FE} - G_{AFE}}{k_B T}\right] = exp\left[\frac{G_0}{k_B T}\epsilon\right], \quad \text{(B.2b)}$$

Hence the y-dependence of the LGD-FSM parameter $\epsilon$ can be estimated from Eqs.(7) and (B.2) and $y > y_{cr}$ as:

$$\epsilon(y) = \frac{k_B T}{G_0} ln[\mu] \approx \frac{k_B T}{G_0} ln\left[C_1\left[exp\left(\frac{y-y_{cr}}{\Delta_1}\right) - 1\right] + C_2\left[1 - exp\left(-\frac{y-y_{cr}}{\Delta_2}\right)\right]\right]. \quad \text{(B.2c)}$$

The **second hypothesis** is that the ceramics can be near the boundary between AFE and FEI phases, and not in a coexisting AFE-FE state. In this case the conditional probabilities of AFE and FEI (indistinguishable from FE by PFM) local states are:

$$W(AFE|AFE \cup FEI) = \frac{exp\left[-\frac{G_{AFE}}{k_B T}\right]}{exp\left[-\frac{G_{AFE}}{k_B T}\right] + exp\left[-\frac{G_{FEI}}{k_B T}\right]}, \quad W(FEI|AFE \cup FE) = \frac{exp\left[-\frac{G_{FEI}}{k_B T}\right]}{exp\left[-\frac{G_{AFE}}{k_B T}\right] + exp\left[-\frac{G_{FEI}}{k_B T}\right]}. \quad \text{(B.3a)}$$

The ratio $\mu$ can regarded equal to the ratio of the AFE and FEI states probabilities calculated theoretically:

$$\mu = \frac{w(AFE|AFE \cup FEI)}{w(FEI|AFE \cup FEI)} = exp\left[\frac{G_{FEI} - G_{AFE}}{k_B T}\right] = exp\left[\frac{G_0}{k_B T}\left(\frac{(1+\epsilon)^2}{4} - \frac{1+\epsilon^2 - \chi(1-\epsilon^2)}{2(1-\chi^2)}\right)\right], \quad \text{(B.3b)}$$

The approximate y-dependence of the parameter $\epsilon$ can be estimated from Eqs.(7) and (10) at small $|\epsilon| \ll 1$ and $y > y_{cr}$ as:

$$\epsilon(y) = 2\frac{k_B T}{G_0}ln[\mu] - \frac{\chi-1}{2(\chi+1)} \approx 2\frac{k_B T}{G_0}ln\left[C_1\left[\exp\left(\frac{y-y_{cr}}{\Delta_1}\right)-1\right] + C_2\left[1 - \exp\left(-\frac{y-y_{cr}}{\Delta_2}\right)\right]\right] - \frac{\chi-1}{2(\chi+1)} \quad (B.3c)$$

The Bayesian formula

$$W(AFE \cup FE|AFE) = \frac{w(AFE|AFE \cup FE)_{W(AFE \cup FE)}}{w(AFE|AFE \cup FE)_{W(AFE \cup E)} + w(AFE|\overline{AFE \cup FE}I)_{W(AFE \cup FEI)}} \quad (B.4)$$

Leads to $W(AFE \cup FE|AFE) = W(AFE \cup FEI|AFE) = \frac{1}{2}$ that in fact is an expected result of a thermodynamic approach we used.